\documentclass[a4paper,12pt]{article}
\usepackage[
        left=3cm,
        right=2cm,
        top=3cm,
        bottom=2cm,
]{geometry}
\usepackage{lipsum}
\usepackage[T1]{fontenc}
\usepackage[utf8]{inputenc}
\usepackage{amssymb}
\usepackage{amsmath}
\usepackage{amsfonts}
\usepackage{authblk}
\usepackage{graphicx}
\usepackage{hyperref}
\usepackage{cite}
\usepackage{setspace}
\usepackage{doi}
\begin{document}
\title{Non-Markovian Effects on Overdamped Systems}
\author[1]{Eduardo dos S. Nascimento\thanks{edusantos18@esp.puc-rio.br}}
\author[1,2]{Welles A. M. Morgado}
\affil[1]{Dept. of Physics, PUC-Rio, Rio de Janeiro, Brazil}
\affil[2]{National Institute for Science and Technology - Complex Systems}

%
%
%
%
\date{} 
%
%
%
\maketitle
\begin{abstract}
We study the consequences of adopting the memory dependent, non-Markovian, physics with the 
memory-less over-damped approximation usually employed to investigate Brownian particles. Due to the finite correlation 
time scale associated with the noise, the stationary behavior of the system is not described by the Boltzmann-Gibbs 
statistics. However, the presence of a very weak external white noise can be 
used to regularize the equilibrium properties. Surprisingly, the coupling to another bath 
effectively restores the dynamical aspects missed by the  over-damped treatment.
\end{abstract}
%
%
%
%
%
%

%
%
\section{Introduction}
Brownian motion is an excellent laboratory for probing the microscopic dynamics of fluid systems. 
For realistic systems, the interaction between the Brownian Particle (BP) and its neighborhood is 
quite complex since, on top of all the microscopic local interactions of the BP and the bath particles, 
the BP's own  motion generates (inertial) hydrodynamic fluxes that interact again with it. In consequence, 
a dissipative memory function arises~\cite{Mori1958,Mori1965}, setting the 
non-Markovian character of the reduced dynamics of the particles. That sort of effect is responsible 
for the anomalies in velocity-velocity correlation functions first pointed out by Alder and Wainwright~\cite{Alder1970}.

It should be clear from the beginning that non-Markovian properties are not properties of 
a system itself, but rather of its description by a chosen set of variables~\cite{livro_vankampen}.
\footnote{As an example, a BP described only by its position will present a non-Markovian behavior. 
On the other hand, by describing it by means of position and velocity will recover the Markovian behavior.} For 
classical systems, the Hamiltonian dynamics, or its equivalent the Liouville dynamics, is 
certainly Markovian. In quantum mechanics the von Neumann equation plays an exactly similar role. 
The existence of a clear cut time-scale separation, such as that due to the mass difference of a 
BP and the surrounding thermal bath, allows us to eliminate (most of) the fast variables of the 
problem via projection operators~\cite{Kampen1985}. This is done by a convenient expansion, of 
the dynamical equations, in powers of a small parameter, such as the ratio of bath-to-BP masses 
\begin{equation*}
    \varepsilon\equiv\sqrt{\frac{m_{bath}}{M_{BP}}}.
\end{equation*}
A didactically interesting example of that procedure 
is described by van Kampen~\cite{Vankampen1986}. In a more general note, powerful methods have been created 
that allow us to carefully separate each phenomena in its correct time-scale, such as the time-extension 
methods~\cite{Sandri1963a,Sandri1963b}. Any remaining slow variables (related to the motion of the BP) are 
subjected to a dynamics that incorporates the effects of the averaged out variables via a non-Markovian dissipative 
term and a rapidly fluctuating noise term. That is the physical justification for the Langevin approach.   
Also, the theory presents a feasible mathematical framework and the stationary properties lead to the well known equilibrium results.

A BP usually exists in  fluid media, at low Reynolds number values (Re), i.e. with a relatively high 
viscosity, where Stokes' Law applies~\cite{livro_landaustatmech}. Stokes' force is proportional to 
product of the viscosity coefficient $\eta$ of the fluid, the velocity $V$ of the BP, and the radius 
$R$ of the BP. The dissipative time scale that arises is proportional to
\begin{equation*}
 \tau_{d} = \frac{M_{BP}}{\gamma}=\frac{\rho_{BP}R^3}{\eta R} = 
 \frac{\rho_{BP}}{\rho}\frac{R}{V}\frac{\rho R V}{\eta} \sim \frac{R}{V}\,\mbox{Re}.
\end{equation*}
A strong approximation is to take the overdamped approximation given by Re $ \rightarrow 0, $ 
as $ M_{BP}/\gamma\rightarrow 0$. In that limit, 
the BP is too light to be able to make  the fluid move by means of its own inertia. 

When the presence of induced flows cannot be ignored, they can be taken into account by a memory kernel 
in the dissipative term of the dynamical equations associated with BP. That kernel is dominated by the 
time-scale $\tau$ associated with the main contribution from the induced slow hydrodynamic modes. 
Without an external source of energy, the BP in contact with a thermal bath will tend to an equilibrium 
state described by the Boltzmann-Gibbs distribution, after a very long transient. 

Short term processes will take the system to an out-of-equilibrium state, unless they are following a 
very specific quasi-static thermodynamic path. However, even for the non-equilibrium cases, it is 
possible to find interesting relations between equilibrium and non-equilibrium  behavoirs~\cite{Jarzynski1997,Crooks2000}. 
Such processes are driven by an external protocol $X(t)$, defined by some time-varying field 
(such as an external force~\cite{Horowitz2007}, a magnetic field~\cite{Batalhao2014}, or the position of the extremity of 
a spring~\cite{Morgado2010}). 

We shall see that using the overdamped approximation, which means assuming a vanishing 
inertial contribution in the Langevin equation, does not lead to an equilibrium distribution 
as $t\rightarrow\infty$. This can be fixed in two ways: either by taking $\tau\rightarrow0$, 
i.e., by assuming Markovian behavior of the BP; or by assuming the presence of an arbitrarily 
weak white noise source. This is enough to reestablish the equilibrium state, no matter how weak the noise source is.

This paper is organized as follows. In section \ref{MassParticle}, we study the massive non-Markovian harmonic bound BP, and 
obtain its equilibrium distribution. Section \ref{OverPartcile} presents the over-damped dynamics and its stationary 
state. In section \ref{Regularizing}, we include a regularizing Gaussian white noise and observe that the equilibrium behavior 
is restored. Finally, the conclusions are stated in section \ref{Conclusions}.

%
%
\section{Massive particle and equilibrium properties} \label{MassParticle}
We initially shall study a 1D Langevin-like model for a massive particle under an 
harmonic external potential as well a Gaussian colored noise. The system is defined by
\begin{equation} \label{LanSys}
m\,\dot{v} +\int_{0}^{t}dt^{\prime}K\left(t - t^{\prime} \right)\, v(t^{\prime})+k\,x = \xi\left( t \right),
\end{equation}
and
\begin{equation}
\dot{x} = v,
\end{equation}
where $K\left(t - t^{\prime} \right)$ is the frictional memory 
kernel~\cite{Mori1958,Mori1965} and $\xi\left( t \right)$ is a stochastic Langevin force. 
The initial conditions are assumed to be
\begin{equation*}
x\left( 0 \right)=0 \quad \text{and} \quad v\left( 0 \right)=0.
\end{equation*}
We have chosen an harmonic external potential for its mathematical convenience, since all 
calculations herein can be made exact and the role of the kernel's time scale be made explicit. 

The Gaussian colored noise is characterized by the cumulants~\footnote{Special care should be taken in order 
to recover the correct limit as $\tau\rightarrow0$. The second cumulant must be $2 
\gamma T \delta(t-t')$, which is assured by the expression in \eqref{NoiseCum}.}
\begin{equation} \label{NoiseCum}
 \begin{split}
  \langle \xi(t)\rangle_c & = 0, \\
  \langle \xi(t) \xi(t^{'})\rangle_c & = \frac{\gamma\,T}{\tau} \exp\left( -\frac{|t-t^{\prime}|}{\tau}\right).
 \end{split}
\end{equation}
where $T$ is the bath temperature and $\tau$ is the correlation time. All other cumulants are 
zero, which means that they do not contribute to the dynamics of the system. Also, due to the noise correlated 
behavior, the memory kernel should of the type
\begin{equation}
    K\left(t - t^{\prime} \right) = \frac{ \langle \xi(t) \xi(t^{'})\rangle_c  }{T},
\end{equation}
in order to be consistent with the fluctuation dissipation theorem.

The approach we are going to use to solve the problem is a straightforward use of the integral 
transformation methods~\cite{Soares-Pinto2006}. We take the Laplace representation of \eqref{LanSys},
\begin{equation}
 \begin{split}
  m\,s\,\tilde{v}(s) & =  -\gamma \frac{\tilde{v}(s)}{(s\,\tau+1)}-k\,\tilde{x}(s)
  +  \tilde{\xi}(s),\\
  s\,\tilde{x}(s) &= \tilde{v}(s),
 \end{split}
\end{equation}
and also for the white noise cumulant~\cite{Morgado2008a},
\begin{equation}
\begin{split}
\langle\tilde{\xi}(z_1)\tilde{\xi}(z_2)\rangle =
  \gamma\,T\,{\frac {2+\tau \, (z_1+z_2)}
  { \left( 1+z_1\tau \right)  \left( 1+z_2\tau \right)
 \left( z_1+z_2 \right) }}. 
\end{split}
\end{equation}
As a result, one can write
\begin{eqnarray*}
\tilde{x}(s) & = & 
\tilde{\xi}(s)\,\frac{(s\,\tau+1)}{R(s)},
\end{eqnarray*}
with
\begin{equation} \label{MassR}
\begin{split}
R(s) &= (m\,s^2+k)\,(s\,\tau+1)+ \gamma\,s \\ 
& =  m\,\tau\,(s-\kappa_1)\,(s-\kappa_2)\,(s-\kappa_3).
\end{split}
\end{equation}
where the roots $\kappa_1,\, \kappa_2,\, \kappa_3$ may obtained exactly in a similar way as 
performed by Soares-Pinto and Morgado \cite{Morgado2008a}.

The model can be solved exactly by means of time-averaging techniques \cite{Morgado2008a,Morgado2010}. 
In fact, the instantaneous probability density associated with \eqref{LanSys} can be written as
\begin{equation}
\begin{split}
P(x,v,t) &= \left( \frac{1}{2\pi} \right)^{2} \int_{-\infty}^{\infty} \int_{-\infty}^{\infty} dQ dP e^{iQx+iPv} \times \\
\\
  & \quad \times \left\langle e^{-iQx\left( t \right) -iPv\left( t \right)} \right\rangle,
\end{split}
\end{equation}
where the moment generating function, as a function of time, can be written as
\begin{equation} \label{MomGen}
 \begin{split}
  \left\langle e^{-iQx\left( t \right) -iPv\left( t \right)} \right\rangle &= \lim_{\epsilon\rightarrow0^+}  
  \sum_{r,n=0}^{\infty}\frac{(-iP)^{r}\,(-iQ)^{n}}{r!n!}\,
  \int_{-\infty}^{\infty}\prod_{a=1}^{n+r}\frac{dq_a}{2\pi}\,
  \exp\left[ \sum_{a=1}^{n+r}(iq_a+\epsilon)\,t\right]\times \\
  &\times \prod_{a=n+1}^{n+r}\left[(iq_a+\epsilon)\,\,
  \frac{((iq_a+\epsilon)\,\tau+1)}{R(iq_a+\epsilon)}\right]\,
  \left<\prod_{a=1}^{n+r}\tilde{\xi}(iq_a+\epsilon)\right>.
 \end{split}
\end{equation}
It is important to bear in mind that $\left\langle \cdot \right\rangle$ is the noise average.
The equilibrium probability distribution is determined by taking the inverse Fourier transform of 
\begin{equation}
 G(Q,P) = \lim_{t \to \infty} \left\langle e^{-iQx\left( t \right) -iPv\left( t \right)} \right\rangle, 
\end{equation}
which is the stationary generating function. 

A simple expression for \eqref{MomGen} is obtained by using the Gaussian properties of the noise, 
which allows us to group the averages into products of averages of pairs. 
The infinite time limit will make the corresponding terms of the form 
\begin{equation}
\exp\left[ \left( iq_a+iq_b+2\epsilon \right)t \right],
\end{equation}
null, unless one integrates over the thermal pole
\begin{equation}
\frac{1}{iq_a+iq_b+2\epsilon}.
\end{equation}
After some straightforward simplifications, the terms $(iq_a+\epsilon)\,\tau+1$ cancel out, 
we obtain products of the integrals of the type
\begin{equation} \label{IntMonGen}
\begin{split}
 I_{xx} &=   \lim_{\epsilon\rightarrow0^+} 
 \int_{-\infty}^{\infty}\frac{dq_1}{2\pi}\, \frac{1}{R(iq_1+\epsilon)R(-iq_1-\epsilon)} \\
 & = -\frac{1}{ 2\,m^2\,\tau^2}\,{\frac {\kappa_{{2}}+\kappa_{{1}}+\kappa_{{3}}}{\kappa_{{1}}\kappa_{{2}}\kappa_{{3}}
 \left( \kappa_{{2}}+\kappa_{{3}} \right)  \left( \kappa_{
{1}}+\kappa_{{3}} \right)  \left( \kappa_{{1}}+\kappa_{{2}} \right)
}} = \frac{1}{ 2\,\gamma\,k},
\\
\\
I_{xv} &=  \lim_{\epsilon\rightarrow0^+} \int_{-\infty}^{\infty}\frac{dq_1}{2\pi}\, 
\frac{iq_1+\epsilon}{R(iq_1+\epsilon)R(-iq_1-\epsilon)} =0,
\\
\\
I_{vv} &=   \lim_{\epsilon\rightarrow0^+}\frac{1}{ m^2\,\tau^2} \int_{-\infty}^{\infty}\frac{dq_1}{2\pi}\,
\frac{(iq_1+\epsilon)^2}{R(iq_1+\epsilon)R(-iq_1-\epsilon)} \\
 & = \frac{1}{2\, m^2\,\tau^2}{\frac {1}{ \left( \kappa_{{2}}+\kappa_{{3}} \right)  \left(
\kappa_{{1}}+\kappa_{{3}} \right)  \left( \kappa_{{1}}+\kappa_{{2}}
 \right) }} = -\frac{1}{ 2\,\gamma\,m}.
\end{split}
\end{equation}
Those integrals allow us to find an analytic expression for the distribution function. In fact, 
since $I_{xv}=0$, no equilibrium average of the type $\left<\tilde{x}\tilde{v}\right>$ 
may exist. Thus, after some algebraic manipulations, one can find
\begin{equation} \label{MonGenEq}
 G(Q,P) = \exp\left[- \frac{T}{2}\left( \frac{Q^2}{k} + \frac{P^2}{m}\right) \right].
\end{equation}
The stationary distribution can be obtained by means of a inverse Fourier transform, which leads to
\begin{equation}\label{pss}
 P^{ss}(x,v) =\frac{\sqrt {km}}{ 2{\pi }T} \exp\left( -{\frac {m{v}^{2}}{2T}}-\frac{k\,x^{2}}{2T} \right) .
\end{equation}
As expected, we find the Boltzmann-Gibbs probability for the equilibrium state of BP
in contact with a thermal reservoir at temperature $T$. This corresponds to the marginal probability 
distribution of the complete system, assuming the interaction energy between the thermal bath particles 
and the BP to be small, where we have averaged out the thermal particles degrees of freedom, which is a 
good approximation for system of hard-sphere type. 

We should observe, from the result above, that the internal dynamics of a system are of no importance with 
respect to its equilibrium distribution. For simple ergodic systems in contact with a thermal reservoir at 
temperature $T$, such as the present one, the final equilibrium state is always given by the Boltzmann-Gibbs 
statistics, regardless whether the equilibration process involves memory or not.

%
%
%
\section{Over-damped case and non-equilibrium distribution} \label{OverPartcile}
We now focus on the fast relaxation dynamics. The idea is to investigate the 
over-damped limit of \eqref{LanSys}, which is obtained by taking the limit $m/\gamma\rightarrow0$. 
Consequently, the Langevin equation is given by
\begin{equation}
     k\,x(t) + \int_{0}^{t}dt^{\prime}K\left(t - t^{\prime} \right)\, 
     \dot{x}(t^{\prime}) =\xi(t).\label{overdampednonmarkovian}
\end{equation}

Another equivalent and simpler way to deal with the model is through the cumulant generating function, 
which is a series expansion the distribution cumulants. Also, due to the gaussian noise properties, only the 
sencond cumulant is relevant to the problem. Then, we have
\begin{equation}
 \begin{split}
  \ln \left\langle e^{-iQx\left( t \right) } \right\rangle &= -
  \frac{Q^2}{2} \lim_{\epsilon \to 0}   
  \int_{-\infty}^{\infty}\frac{dq_1 e^{(iq_1 +\epsilon)\,t }}{2\pi} 
   \times \\ 
  & \quad \times \int_{-\infty}^{\infty} \frac{dq_2 e^{(iq_2 +\epsilon)\,t }}{2\pi} \times \\
  & \quad \times \left< \tilde{x}(iq_1+\epsilon) \tilde{x}(iq_2+\epsilon)\right>_{c} .
 \end{split}
\end{equation}
where
\begin{equation} \label{OverLaplace}
  \tilde{x}(s)= \frac{s\,\tau+1}{(\gamma+k\,\tau)s + k} \, \tilde{\xi}(s),
\end{equation}
is calculated through the Laplace transform of \eqref{overdampednonmarkovian}. As a result, it is not difficult to show that
\begin{equation}
 \begin{split}
  \ln \left\langle e^{-iQx\left( t \right) } \right\rangle &= -\frac{Q^2}{2}\frac{\gamma T}{k\left( k\tau + \gamma \right)} \times \\
   & \quad \times \left[ 1 - \left( 1-\tau a \right)
    e^{ -2at } \right],
 \end{split}
\end{equation}
where
\begin{equation}
    a = \frac{k}{k\tau + \gamma}.
\end{equation}
For $t \to \infty$, the stationary cumulant generating function is given by 
\begin{equation*}
\ln G(Q) = -\frac{T\,\gamma\,Q^2}{2k\,(\gamma+k\,\tau)},
\end{equation*}
which allows us to write the probability distribution 
\begin{equation} \label{pssoverdamped}
 \begin{split}
  P^{ss}(x) 
  &=  \int_{-\infty}^{\infty}\frac{dQ}{2\pi} \, e^{iQ\,x} \,
  \exp\left[-\frac{T\,\gamma\,Q^2}{2k\,(\gamma+k\,\tau)}\right], \\
  \\
  &=  \sqrt{\frac{k\,(\gamma+k\,\tau)}{2\pi\,T\,\gamma}} \,
  \exp\left[-\frac{k\,(\gamma+k\,\tau)\,x^2}{2\,T\,\gamma}\right].
 \end{split}
\end{equation}
Cleary, the stationary state is not described by a Boltzmann-Gibbs distribution.

Note that \eqref{pssoverdamped} leads to a renormalization of the temperature 
felt by the particle,
\begin{equation} 
    T_{eff}\equiv \frac{T\,\gamma}{\gamma+k\,\tau}.\label{teff}
\end{equation}
Clearly, this is unphysical since we have seen that the presence of mass 
yields the correct physical limit. That result had  
already been obtained earlier by Cugliandolo and Kurchan~\cite{Cugliandolo2000} and studied in detailed by Soares-Pinto and Morgado\cite{Morgado2008a}.

The physical reason for the discrepancy is that one of the main roles of a memory kernel is to be the witness to slow modes, 
such as fluid streamlines, generated in the bath by its interaction with the BP. These flows only appear due to the inertia of the BP that is capable of moving the masses of the particles of the bath out of 
its way. On the other hand, the over-damped approximation assumes exactly the opposite: the BP inertia's is irrelevant! In fact, the relaxation of the BP is supposed to be so fast that it assumes that the BP follows the instantaneous local velocity of the fluid of bath particles. From it we see that by taking $\tau\rightarrow0$ makes the whole analysis consistent, and \eqref{teff} becomes 
$T_{eff}= T$, as equilibrium requires.

This is due to artificially mixing the overdamped, memory-less, model with the  physics of a non-Markovian  model exhibiting a memory function. 
It is important to bear in mind that our result is consistent with the analysis present by Soares-Pinto and Morgado~\cite{Morgado2008a}. Therein the authors shown that, 
in order to recover the correct equilibrium state, another white Gaussian 
noise (internal or external) should be included and with same temperature of the non-Markovian noise. 
Clearly, the usual equilibrium might also be obtained by taking the limit $\tau\rightarrow0$ above. 

From a mathematical perspective, the presence of mass (independently of its value) implies that $R(s)=0$ (see \eqref{MassR}) is a cubic equation with three roots. However, the over-damped limit leads to a simple linear solution, as shown in \eqref{OverLaplace}. This suggests that, as $m\rightarrow0$, 
two of the cubic roots must move to $x\rightarrow\infty.$ The solutions will be quite different.

Let us analyze the problem as $m/\gamma\rightarrow0$. In this limit, it is straightforward to show that the roots presents the asymptotic behavior
%
\begin{equation}
 \begin{split}
  \kappa_1 &\rightarrow -\frac{k}{\gamma+k\,\tau}, \\
  \kappa_2 &\rightarrow -\frac{\gamma}{2\tau(\gamma+k\,\tau)} +\mathrm{i}\,\sqrt{\frac{\gamma+k\,\tau}{m\,\tau}}, \\
  \kappa_3 &\rightarrow -\frac{\gamma}{2\tau(\gamma+k\,\tau)} -\mathrm{i}\,\sqrt{\frac{\gamma+k\,\tau}{m\,\tau}},
\end{split}\label{asympt}
\end{equation}
which can easily be checked out numerically. As a result, the over-damped limit leads to a finite value for $\kappa_1$, although the other roots tend to $-c\pm \mathrm{i}\,\infty$. By taking first the limit $m\rightarrow0$ in \eqref{LanSys}, which possesses three dynamical poles in Fourier-Laplace representation, we directly obtain 
\eqref{overdampednonmarkovian}, which, from \eqref{asympt}, presents just a single dynamical pole. It is clear that 
this is a consequence of taking the mass-less limit, before solving \eqref{LanSys}, 
because it makes the other two roots move to infinity, and they no longer participate on the dynamics. 
Next we see how to regularize that behavior and reach equilibrium once again while still keeping 
the over-damped non-Markovian character.

The result \eqref{teff} reminds us of the temperature slip between the interface of a gas 
and a limiting wall, such as the gas-wall interaction model described in chapter 6 of 
Chapman \& Cowlings's book~\cite{livro_chapman}. That slip arises because the particles 
that are momentarily adsorbed in the wall will be expelled back into the gas, but they will 
not collide before moving by a typical length of the order of the mean-free-path, reflecting 
a decrease in the equilibration efficiency near the wall. Despite the fact that the problem at hand is quite 
distinct, it is clear that, by neglecting the mass of the particle, the mobility of the BP is 
clearly reduced and the typical distance it will hover around the origin of the coordinates is reduced by a factor
\begin{equation}
\sqrt{ \frac{\gamma}{(\gamma+k\tau)} }.
\end{equation}

\section{Regularizing the equilibration via an additional noise} \label{Regularizing}
Regularization of the overdamped non-Markovian process can be achieved by replacing the 
missing kinetic energy of the BP via other means, i.e., an external source such as a 
thermal bath. That extra noise must be weak enough so that the energy transfer from it 
to the BP is actually negligible compared with the that due to the main non-Markovian 
noise. In fact its role will be to give back the missing particle's ``mobility'' due 
to the extreme rate of damping.

In order to see how this work, we replace \eqref{overdampednonmarkovian} by
\begin{equation} \label{overdampednonmarkovianregularized}
 \begin{split}
     k\,x(t) &+ \int_{0}^{t}dt^{\prime}K\left(t - t^{\prime} \right)\, \dot{x}(t^{\prime}) + \\
     & + \Gamma\, \dot{x}\left( t \right) = \eta(t)+\xi(t),
 \end{split}
\end{equation}
where a second Gaussian noise is included,
\begin{equation} \label{NewNoise}
   \left\langle \eta(t)\,\eta(t') \right\rangle = 2\Gamma T^{\prime} \delta(t-t'), 
\end{equation}
with zero average and bath temperature $T^{\prime}$.
Consequently, the transformed Langevin equation can be rewritten as 
\begin{equation}
\tilde{x}(s)  = \frac{(s\,\tau+1)}{R_{1}(s)}\,\left[\tilde{\eta}(s) +  \tilde{\xi}(s)\right],
\end{equation}
where
\begin{equation}
    R_{1}(s)= \Gamma\tau(s-\lambda_{+})
    (s-\lambda_{-}),
\end{equation}
with
\begin{equation}
\lambda_{\pm} = - {\frac {k\tau+\gamma+\Gamma \pm\sqrt {\tau\,k \left( k\tau+2\,
\gamma-2\,\Gamma  \right) + \left( \gamma+\Gamma  \right) ^{2}}}{2\,\Gamma \,\tau}}.
\end{equation}
The Fourier-Laplace representation of the noise cumulant \eqref{NewNoise} is given by
\begin{equation}
 \left\langle \tilde{\eta}\left( s_1 \right)\tilde{\eta}\left( s_2 \right)  \right\rangle
  = \frac{2\Gamma T^{\prime}}{s_1+s_2}.
\end{equation}
Using the time-averaging treatment, one can find that the system presents a stationary 
behavior characterized by the cumulant generating function 
\begin{equation} \label{2Noise}
 \begin{split}
  \ln G(Q) &= -Q^2 \,\lim_{\epsilon\rightarrow0} \int_{-\infty}^{\infty} 
  \frac{dq_1}{2\pi} \,\frac{1 }
  {R(iq_1+\epsilon)} \times \\ 
  & \times \frac{1 }{R(-iq_1-\epsilon)} 
  \left[  \gamma\, T+ \Gamma\, T^{\prime}\, 
  \times \right. \\ 
  & \left. \times \left( 1+(iq_1+\epsilon)\tau \right) \left( -(iq_1+\epsilon)\tau+1 \right)\right] \\
   &= -\frac{Q^2}{2}\,{\frac {T^{\prime}\tau\,k+ \Gamma \,T^{{\prime}}+
  \gamma\,T}{k \left( k\tau+\gamma+\Gamma  \right) }},
 \end{split}
\end{equation}
which is consistent with the overdamped limit already obtained 
by Soares-Pinto and Morgado \cite{Morgado2008a}. 

Now, assuming that the two baths have the same temperature, $T^{\prime}=T$, the Gibbsian state is reestablished,
\begin{equation}
 \begin{split}
  \ln G(Q) &= -\frac{Q^2}{2}\,\frac{T}{k}, \\
  \\
  P_{eq}(x) &=\sqrt{\frac{k}{2\pi T}}\,\exp\left(-\frac{k\,x^2}{2T} \right),
 \end{split}
\end{equation}
independently of the value of $\Gamma$. Note that, as long as $\Gamma$ is small, but not zero, 
the two dynamical poles $\lambda_{\pm}$ are still integrated over, and equilibrium is 
reached. Also, it should be noticed that, for the above, the dissipation and injection of 
energy due to the additional white noise are negligible. However, the correct 
equilibration is obtained with (almost) cost free as a true regularizing procedure should be.

In fact, the rates of energy transfer and dissipation go to zero for vanishing values of 
$\Gamma$, although the missing average kinetic energy is restored at equilibrium.

%
%
\section{Conclusions} \label{Conclusions}
Many physical phenomena are fundamentally influenced by their typical characteristics time-scales. 
Indeed, the paradigmatic BP model can be viewed basically as a time-scale separation problem: 
fast degrees of freedom, associated with the fast thermal bath particles, can be averaged out, 
yielding the much slower dynamics associated with the BP. In that context, the Langevin equation 
effectively describes the time-scales associated with the mechanical oscillations, 
the dissipation and the memory or persistence. 

The overdamped model assumes that the BP moves in an extremely viscous fluid, where dissipation 
is very fast. For that to happen it is necessary that $m/\gamma\rightarrow0$, which does not 
unequivocally imply that $m\rightarrow0$ since taking $\gamma\rightarrow\infty$ would also do. However, 
one of the most used forms of the overdamped approximation is to simply take the inertial term 
$m\ddot{x}$ to zero on the Langevin equation. Such a  choice lead to a much rougher trajectory 
for the BP when compared with the massive case~\cite{Sekimoto2010}.

We have seen that another consequence of the simple choice for the overdamped approximation, 
in the context of non-Markovian dynamics, is that the stationary state will not correspond to 
the Boltzmann-Gibbs form of the equilibrium distribution. This happens because taking $m\rightarrow0$ 
in the dynamics means to make two dynamical poles move to infinity where they will not contribute to 
shaping the equilibrium state. In fact, any nonzero value for the mass, no matter how small, would lead 
to the correct equilibrium distribution at long times. 

It is still possible to regularize the overdamped non-Markovian behavior through an additional 
(white) Gaussian noise with a dissipative factor arbitrarily small. The inclusion of an extra thermal bath guarantees 
that the final equilibrium state will arise despite the presence of an extremely weak uncorrelated noise.

\section*{Acknowledgement}
We thank the financial support of the the Brazilian agency CAPES.


\begin{thebibliography}{99}

\bibitem{Mori1958}
H.~Mori.
\newblock {Statistical-Mechanical Theory of Transport in Fluids}.
\newblock {\em Physical Review}, 112(6):1829, 1958.

\bibitem{Mori1965}
H~Mori.
\newblock {Transport, Collective Motion, and Brownian Motion}.
\newblock {\em Prog. Theor. Phys.}, 33(06):423, June 1965.

\bibitem{Alder1970}
B.~J. Alder and T.~E. Wainwright.
\newblock {Decay of the velocity autocorrelation function}.
\newblock {\em Physical Review A}, 1(1):18--21, 1970.

\bibitem{livro_vankampen}
N.~G. van Kampen.
\newblock {\em Stochastic Processes in Physics and Chemistry}.
\newblock North-Holland, Amsterdam, 1992.

\bibitem{Kampen1985}
N.~G. van Kampen.
\newblock {Elimination of fast variables}.
\newblock {\em Physics Reports}, 124(2):69, dec 1985.

\bibitem{Vankampen1986}
N.~G. van Kampen and I~Oppenheim.
\newblock {Brownian motion as a problem of eliminating fast variables}.
\newblock {\em Physica A: Statistical and Theoretical Physics},
  138(1-2):231--248, September 1986.

\bibitem{Sandri1963a}
G.~{Sandri}.
\newblock {The foundations of nonequilibrium statistical mechanics, II}.
\newblock {\em Annals of Physics}, 24:380, 1963.

\bibitem{Sandri1963b}
G.~{Sandri}.
\newblock {The foundations of nonequilibrium statistical mechanics, I}.
\newblock {\em Annals of Physics}, 24:332, 1963.

\bibitem{livro_landaustatmech}
L.D. Landau and E.M. Lifshitz.
\newblock {\em Statistical Mechanics}.
\newblock Elsevier, Amsterdam, 1980.

\bibitem{Jarzynski1997}
C~Jarzynski.
\newblock {Equilibrium free-energy differences from nonequilibrium
  measurements: A master-equation approach}.
\newblock {\em Physical Review E}, 56(5):5018--5035, 1997.

\bibitem{Crooks2000}
G.~E. {Crooks}.
\newblock {Path-ensemble averages in systems driven far from equilibrium}.
\newblock {\em {Physical Review E}}, {61}:{2361}, {2000}.

\bibitem{Horowitz2007}
Jordan Horowitz and Christopher Jarzynski.
\newblock {Comparison of work fluctuation relations}.
\newblock {\em Journal of Statistical Mechanics: Theory and
  Experimentechanics}, 11:P11002, 2007.

\bibitem{Batalhao2014}
Tiago~B. Batalh{\~{a}}o, Alexandre~M. Souza, Laura Mazzola, Ruben Auccaise,
  Roberto~S. Sarthour, Ivan~S. Oliveira, John Goold, Gabriele {De Chiara},
  Mauro Paternostro, and Roberto~M. Serra.
\newblock {Experimental reconstruction of work distribution and study of
  fluctuation relations in a closed quantum system}.
\newblock {\em Physical Review Letters}, 113(14):1--5, 2014.

\bibitem{Morgado2010}
W~A~M Morgado and D~O Soares-Pinto.
\newblock {Exact nonequilibrium work generating function for a small classical
  system}.
\newblock {\em Physical Review E}, 82:021112, 2010.

\bibitem{Soares-Pinto2006}
D~O Soares-Pinto and W~A~M Morgado.
\newblock {Brownian dynamics , time-averaging and colored noise}.
\newblock {\em Physica A}, 365:289--299, 2006.

\bibitem{Morgado2008a}
Diogo~Oliveira Soares-Pinto and W~A~M Morgado.
\newblock {Exact time-average distribution for a stationary non-Markovian
  massive Brownian particle coupled to two heat baths}.
\newblock {\em Physical Review E}, 77:011103, 2008.

\bibitem{Cugliandolo2000}
L.~F. Cugliandolo and J.~Kurchan.
\newblock {A scenario for the dynamics in the small entropy production limit}.
\newblock {\em Journal of the Physical Society of Japan}, 69:247--256, 2000.

\bibitem{livro_chapman}
S.~Chapman and T.G. Cowling.
\newblock {\em The Mathematical Theory of Non-Uniform Gases}, chapter The
  non-uniform state for a simple gas.
\newblock Cambridge University Press, Cambridge, 1970.

\bibitem{Sekimoto2010}
Ken Sekimoto.
\newblock {\em {Stochastic Energetics}}, volume 799 of {\em Lecture Notes in
  Physics}.
\newblock Springer Berlin Heidelberg, Berlin, Heidelberg, 2010.

\end{thebibliography}
\end{document}